\documentclass[manuscript]{aastex}

\usepackage{graphicx}
\usepackage{color}

\begin{document}

\title{Broadening of H$_2$O rotational lines by collisions with He atoms at low temperature}
\author{M. I. Hern\'andez}
\affil{Instituto de F\'isica Fundamental CSIC}
\affil{Serrano 123, 28006 Madrid, Spain}
\and
\author{J. M. Fern\'andez, G. Tejeda, E. Moreno, and S. Montero }
\affil{Instituto de Estructura de la Materia CSIC}
\affil{Serrano 121, 28006 Madrid, Spain}
\email{jm.fernandez@csic.es}

\date{\today}


\begin{abstract}
We report pressure broadening coefficients for the 21 electric-dipole transitions between the eight lowest
rotational levels of ortho-H$_2$O and para-H$_2$O molecules by collisions with He at temperatures from 20 to 120 K.
These coefficients are derived from recently published experimental state-to-state rate coefficients for H$_2$O:He inelastic collisions, plus an elastic contribution from close coupling calculations.
The resulting coefficients are compared to the available experimental data.
Mostly due to the elastic contribution, the pressure broadening coefficients differ much from line to line, and increase markedly at low temperature.
The present results are meant as a guide for future experiments and astrophysical observations.
\end{abstract}


\maketitle

\section{Introduction}
                     
Pressure broadening (PB) of spectral lines is useful for remote sensing in Astrophysics,
allowing for a straightforward determination of the number density of colliders.
Since H$_2$O is a relevant observational target in present day Astrophysics \citep{dish13},
we focus the discussion below onto this species, in particular on the 
broadening induced by collisions with He atoms.
Microwave measurements of pressure broadening coefficients (PB-coefficients in short)
of rotational lines of H$_2$O by collisions with He at low temperature have been reported by \cite{goye90,dutt93,dick10}.
These experiments pose, however, severe difficulties due to the strong tendency of gaseous 
water to condense, and are limited to the few spectral lines within reach of the particular instrument.

On the other hand, PB-coefficients are often derived from the state-to-state rate coefficients (sts-rates in short) for inelastic collisions. 
We have recently published \citep{teje15} a laboratory study of the rotational inelastic collisions of H$_2$O with He atoms at low temperature (20--120~K).
In that work a new collection of experimentally-derived sts-rates was reported, with 1$\sigma$ uncertainty of $\approx 6$\% at 120~K and $\approx 11$\% at 20~K.

There is, however, a further contribution to the PB-coefficients due to interferences between elastic amplitudes, which can be relevant in some cases. Here, we have obtained this elastic term from close-coupling (CC) calculations, employing the potential energy surface (PES) by \cite{patk02}. The relative importance of the elastic and inelastic contributions to the broadening of H$_2$O spectral lines by helium is assessed. 

The main goal of the present work is to provide an extended set of PB-coefficients for H$_2$O lines by collisions with He, based on the experimental sts-rates. This set includes most transitions observed in the Water In Star-forming regions with Herschel (WISH) program \citep{dish11}. 
These PB-coefficients, which are a sum of inelastic and elastic contributions from experiments and theory, respectively, allow us for a revision of the PB-coefficients published so far. 

\section{Theoretical frame for the pressure broadening of spectral lines}

At densities low enough to prevent strong line mixing, the Lorentzian half-width at half-maximum $\Gamma_{i f}$ of an $i \rightarrow f$ spectral line is given by 
\begin{equation}
   \Gamma_{i f} = \gamma_{i f} \, p,                               \label{Gammap}
\end{equation}
where $p$ is the total pressure of colliders and  $\gamma_{i f}$ is the pressure broadening coefficient, which depends only on the translational temperature $T$ of the bath and on the nature of the colliders;
$\gamma_{i f}$ can be expressed in terms of a thermally averaged pressure broadening cross section $\sigma_{i f}(T)$ (PBx-sections in short) as
\begin{equation}
  \gamma_{i f}(T)= \frac{1}{2\pi} \frac{1}{k_B T}  \, v \, \sigma_{i f}(T) ,              \label{gammasigma}
\end{equation}
where $k_B$ is Boltzmann constant,  and $  v = \sqrt{ (8 k_B T)/(\pi \mu)}$ is the mean relative velocity of colliding partners of a reduced mass $\mu$.

The PBx-section is usually given as a sum of two contributions \citep{bara58,wies10}
\begin{equation}
  \sigma_{i f}(T)= \sigma^{(in)}_{i f}(T) + \sigma^{(el)}_{i f}(T) ,              \label{sigma in-el}
\end{equation}
one inelastic, $\sigma^{(in)}_{i f}(T)$, and another elastic, $\sigma^{(el)}_{i f}(T)$.
The inelastic contribution 
can be calculated in favorable cases from the sts-rates $ k_{i \rightarrow j}(T)$ for inelastic collisions, tabulated in databases like BASECOL \citep{dube13} or LAMDA \citep{scho05}, by means of
\begin{equation}
  \sigma^{(in)}_{i f}(T) =  \frac{1}{2  v } 
    \left [ \sum_{j \neq i} k_{i \rightarrow j}(T)  + \sum_{j \neq f} k_{f \rightarrow j}(T) \right ].              \label{sigmaks}
\end{equation}
On the other hand, the elastic contribution, sometimes referred to as ``dephasing'' \citep{thib00}, 
\begin{equation}
   \sigma^{(el)}_{if} (T)= \left  \langle \int d\Omega \mid f_i(\Omega,E_k) -   f_f(\Omega,E_k) \mid^2 \right \rangle_T ,
\label{def-contrib-e}
\end{equation}
is due to the interference between the elastic scattering amplitudes $f_i$ and $f_f$ for the two states involved in the transition \citep{bara58,wies10},
which are functions of the scattering angle, $\Omega$, and of the kinetic energy, $E_k$, while $\langle \; \rangle_T$ indicates a thermal average.
This contribution $\sigma^{(el)}_{if} (T)$ cannot be derived straightforwardly neither from tabulated material nor from experiment. However, it can be calculated through advanced quantum methods like the CC approach, along with a good PES for the colliding pair.

The elastic contribution $\sigma^{(el)}_{i f}(T)$ to the PBx-section vanishes for the isotropic part of the Q-branch ($\Delta J=0)$ lines in the vibrational Raman spectra.
On this ground, it has often been assumed that $\sigma^{(el)}_{i f}(T)$ is considerably smaller \citep{gree90}, or even negligible \citep{dick10}, than the inelastic contribution $\sigma^{(in)}_{i f}(T)$ for the 
electric-dipole or quadrupole absorption/emission lines in the infrared and microwave regions. 
This can be explained in part  because of the convenience of  Eq.~(\ref{sigmaks}) and the availability of the required data, and in part because it has been proven so for a number of small molecules  \citep{gree80, palm86,gree89,thib00,thib02,thib09}.
For asymmetric top molecules, however, the elastic contribution has been shown to be important at least for the pressure broadening of H$_2$O lines by H$_2$ \citep{wies10,drou12}.
The extent of elastic and inelastic contributions to the pressure broadening of H$_2$O lines by He is discussed below.

\section{Procedure and results}

\begin{table}[t]
 \caption{Identification of rotational energy levels of H$_2^{16}$O in the vibrational ground state, after \cite{tenn01}.}
 \begin{center}
 \begin{tabular}{rrrrr|rrrrr} \hline  \hline
   \multicolumn{5}{c}{para-H$_2$O} & \multicolumn{5}{c}{ortho-H$_2$O}  \\
 $i$ & $E_i$ (cm$^{-1}$) & $J$ & $K_a$ & $K_c$ & $i$ & $E_i$ (cm$^{-1}$) & $J$ & $K_a$ & $K_c$ \\ \hline    
  1  &      0.0000       &  0  &   0   &   0   &  1  &      23.7944      &  1  &   0   &   1   \\           
  2  &     37.1371       &  1  &   1   &   1   &  2  &      42.3717      &  1  &   1   &   0   \\           
  3  &     70.0908       &  2  &   0   &   2   &  3  &      79.4964      &  2  &   1   &   2   \\           
  4  &     95.1759       &  2  &   1   &   1   &  4  &     134.9016      &  2  &   2   &   1   \\           
  5  &    136.1639       &  2  &   2   &   0   &  5  &     136.7617      &  3  &   0   &   3   \\           
  6  &    142.2785       &  3  &   1   &   3   &  6  &     173.3658      &  3  &   1   &   2   \\           
  7  &    206.3014       &  3  &   2   &   2   &  7  &     212.1564      &  3  &   2   &   1   \\           
  8  &    222.0527       &  4  &   0   &   4   &  8  &     224.8384      &  4  &   1   &   4   \\           
 \hline
  \end{tabular}
  \end{center}
\end{table}

\begin{figure}[t]
  \includegraphics*[width=0.8\textwidth]{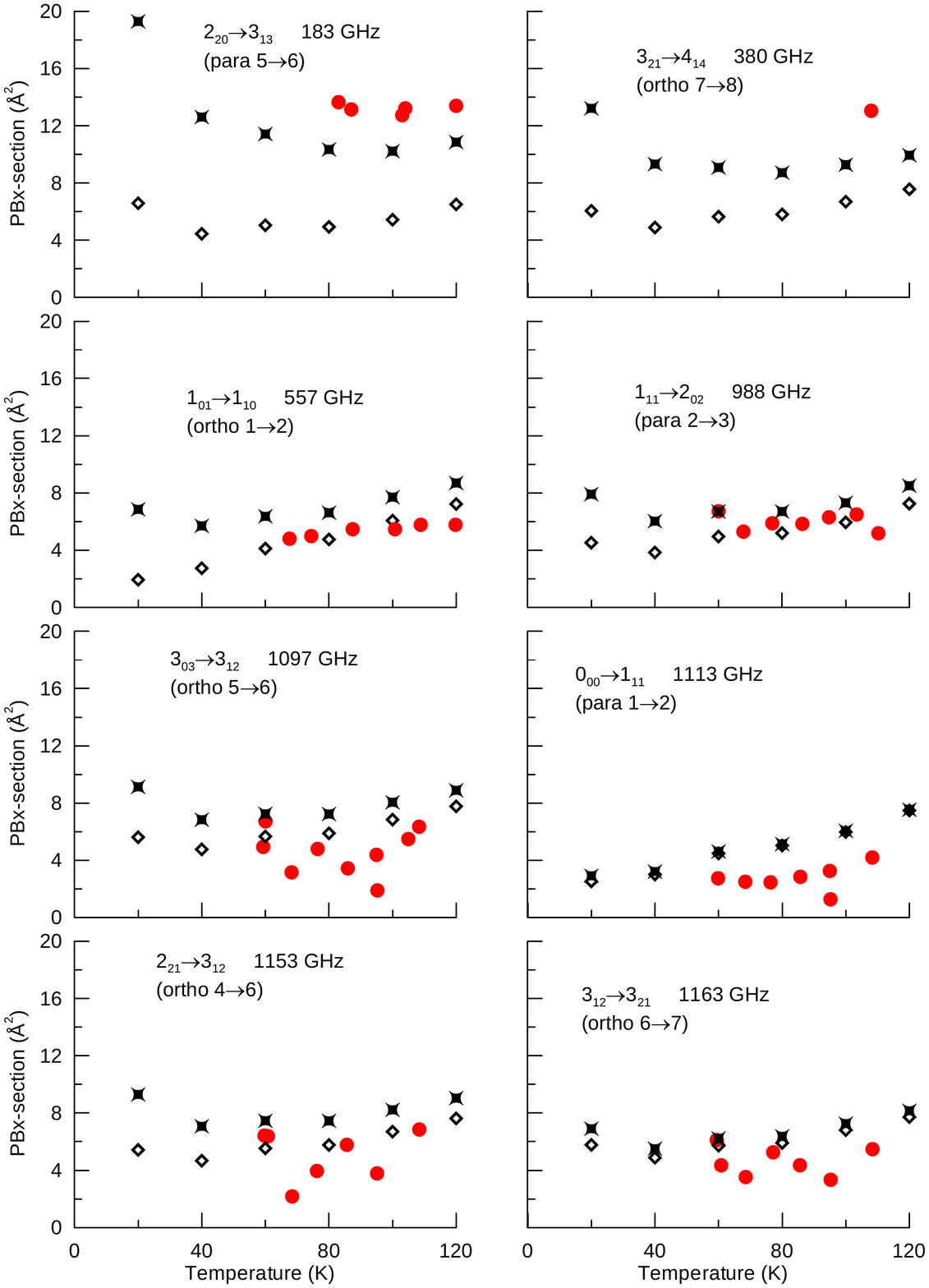}
  \caption{Experimental (bullets), inelastic (open rhombs) and total (stars) PBx-sections for some rotational lines of H$_2$O by collisions with He.}
\end{figure}

Labels $i$ for the rotational energy levels of para-H$_2$O and ortho-H$_2$O relevant for the present work are given in  Table~1.
Experimental PB-coefficients $\gamma_{i f}$ for just eight rotational lines of H$_2$O have been reported so far by \cite{goye90,dutt93,dick10}
for collisions with He at low temperature.
For an easier comparison with the calculations, it is convenient to transform them into the corresponding PBx-sections by means of Eq.~(\ref{gammasigma}) expressed as
\begin{equation}
  \sigma_{i f}= 0.4472 \, \sqrt{\mu T}  \,  \gamma_{i f} ,
\end{equation}
where $\sigma_{i f}$ is in \AA$^2$ for $\gamma_{i f}$ given in MHz/Torr, $\mu$ in amu, and $T$ in K.
The above quoted experimental data are plotted in Fig.~1 as bullet symbols,
where the data from \cite{dick10} are referred to the temperature calibration $T_{\rm gas}=0.8849 \, T_{\rm cell}+ 41.62$.
A preliminary interpretation of the PB-data by \cite{dick10} was attempted by these authors on the basis of Eq.~(\ref{sigmaks}), i. e., neglecting the elastic contribution to the PBx-section, and using the 
sts-rates calculated by \cite{gree93}. These sts-rates have been shown to be significantly smaller than the experimental ones by \cite{teje15}, and also smaller than those calculated by \cite{yang13} employing the improved H$_2$O-He PES by \cite{patk02}.
Although \cite{dick10} attained a reasonable agreement between theory and experiment for the six lines over 500 GHz, we noticed that such interpretation, neglecting $\sigma^{(el)}_{i f}$, sharply disagrees  
for the $5 \rightarrow 6$ (183~GHz) line of para-H$_2$O by \cite{goye90}   
and for the $7 \rightarrow 8$ (380~GHz) line of ortho-H$_2$O by \cite{dutt93}
(see Fig.~1, top panels).
This led us to suspect that the elastic contribution might be significant for the pressure broadening of H$_2$O lines by He.

In order to confirm or to refute the above conjecture, we have proceeded as follows.
First, referring to Eq.~(\ref{sigma in-el}), we have calculated $\sigma^{(in)}_{i f}(T)$ for the H$_2$O lines according to Eq.~(\ref{sigmaks}) using the experimental sts-rates for H$_2$O:He inelastic collisions by \cite{teje15}. These $\sigma^{(in)}_{i f}(T)$ are plotted in Fig.~1 as open rhombs. Then, for the elastic contribution  $\sigma^{(el){\rm CC}}_{i f}(T)$ 
 we have carried out CC calculations with the MOLSCAT code \citep{molscat} based on the PES by  \cite{patk02}, which provides the best agreement with the experimental sts-rates for inelastic collisions \citep{teje15}.
Details about the calculations are given in Appendix A.
 
The total PBx-sections, plotted in Fig.~1 as black stars, are a sum
\begin{equation}
  \sigma^{{\rm TOTAL}}_{i f}(T) = \sigma^{(in)}_{i f}(T) +  \sigma^{(el){\rm CC}}_{i f}(T) ,
\end{equation}
 of an experimental inelastic term plus a calculated elastic one.
They show a better agreement with the apparently anomalous experimental PBx-sections of the $5 \rightarrow 6$ (183~GHz) line of para-H$_2$O  
and the  $7 \rightarrow 8$ (380~GHz) line of ortho-H$_2$O.
The difference (stars minus open rhombs) clearly shows that, at low temperature, the elastic contribution to the total PBx-section is larger than the inelastic one for these two lines.

In view of the above results, we have extended the described procedure to a number of rotational lines suitable for astrophysical diagnostics of  H$_2$O densities in media dominated by collisions with helium.
PBx-sections  $\sigma_{i f}$ and coefficients $\gamma_{i f}$,  both inelastic-only and total, for the 21 microwave lines  between the eight lowest rotational levels of para-H$_2$O and ortho-H$_2$O due to collisions with He, are reported in Table~2 for six temperatures between 20 and 120~K.

\section{Discussion}

First we discuss the inelastic contribution, then the elastic one, and finally the total PBx-sections of H$_2$O by helium in the 20-120~K thermal range.

The {\em inelastic} contribution to the thermally averaged PBx-section of most lines reported in Fig.~1 and Table~2 show a similar pattern. They range from 4 to 8~\AA$^2$ for temperatures $ 20 \leq T \leq 120$~K, showing a shallow  minimum at $T \simeq 40$~K, and lying, for a given $T$, within $\pm 1$~\AA$^2$ from the average, reaching an almost constant value of 7.4~\AA$^2$ at 120~K.
Exceptions to this pattern are the lines at 557 and 1113~GHz involving the lowest energy levels ($1 \rightarrow 2$)  of para-H$_2$O and ortho-H$_2$O, whose inelastic contribution to the PBx-section decreases monotonically with the temperature down to $\sigma^{(in)}_{i f} < 3$~\AA$^2$ at 20~K.
This behavior can be rationalized as a statistical effect, since the number of available inelastic excitation and relaxation channels increases much with rotational energy for an asymmetric top molecule, and thus the summations of sts-rates in Eq.~(\ref{sigmaks}) tend to average.
The two exceptions are the ($1 \rightarrow 2$) lines involving the ground states of each species, which can only be inelastically excited, in addition to the lower density of available states.

\begin{figure}[t]
  \includegraphics*[width=1.0\textwidth]{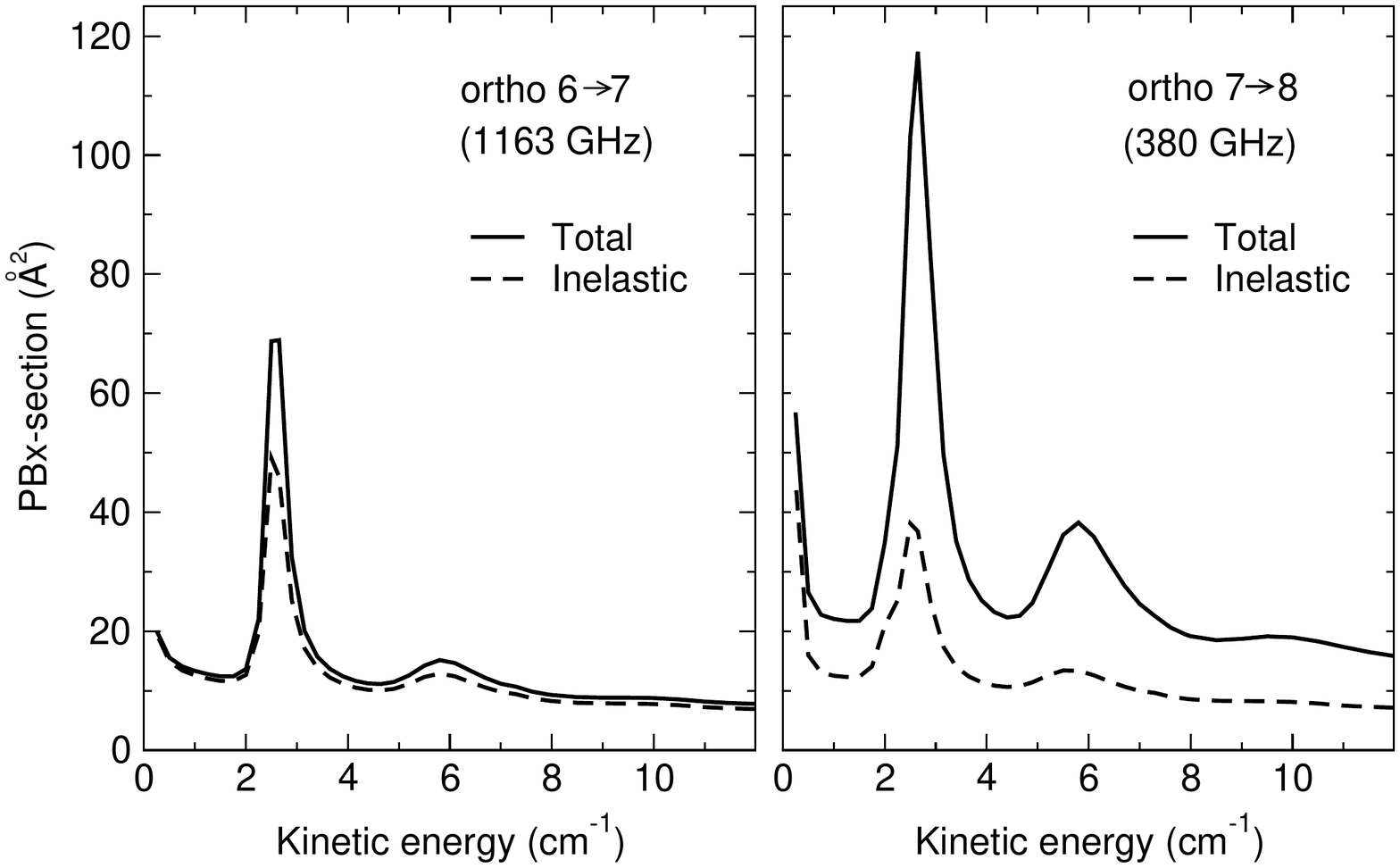}
  \caption{PBx-sections versus kinetic energy for two rotational lines of H$_2$O.}
\end{figure}

The {\em elastic} contribution, in contrast, depends more specifically on the particular $i \rightarrow f$ transition. In addition, this component always increases at lower temperatures, and can be significantly large at 20~K, causing
a considerable dispersion in the {\em total} PB-coefficients.
The described dependence with temperature had already been observed in linear molecules \citep{thib00,thib01,thib02}, 
where it was also noted that the elastic contribution is larger for lower values of the rotational quantum number $J$, in exact correlation with the rotational energy and with the line frequency. 
In the present case, only a coarse inverse correlation of the elastic contribution with the line frequency can be guessed, and, due to the asymmetric top character of H$_2$O, there is no simple relation between line frequencies and rotational energy or angular momentum.
For example, the lines at 1153~GHz and 1163~GHz are close in frequency but they show different elastic PB, while the lines at 380~GHz and 1661~GHz, far in frequency, have similar elastic PB.
This behavior is further confirmed by other lines in Table 2. 

We present in Fig.~2 the dependence with the kinetic energy of the total and inelastic PBx-sections for two representative lines, the $6 \rightarrow 7$ (1163~GHz)  and  the $7 \rightarrow 8$ (380~GHz) lines of ortho-H$_2$O. Both transitions exhibit some low-energy resonances (at $\sim 2.5$ and $\sim 6$~cm$^{-1}$), already reported as shape resonances by \cite{dagd10}.
For the  6 $\rightarrow$ 7 line, the elastic contribution (inspected by difference between the total and the inelastic  PBx-sections) is negligible but at the resonances. On the contrary, for the 7 $\rightarrow$ 8 line, this contribution is significant for all energies, and increases near the resonances.
Enhancement of the elastic contribution at resonances can be explained by sharp changes in the scattering amplitudes as functions of the kinetic energy.
These resonances, which appear for all the transitions studied here, play an increasing role in the thermally averaged PBx-sections as temperature decreases.

Regarding the {\em tota}l PBx-sections, as temperature decreases most of them drop, reach a minimum at different $T$ depending on the line, and then grow up. The exception is, again, the $1 \rightarrow 2$ (1113 GHz) line of para-H$_2$O, which decreases monotonically in the 120 to 20~K thermal range.
Calculated total PBx-sections by \cite{malu92} for the $5 \rightarrow 6$ (183~GHz) line of para-H$_2$O and the $7 \rightarrow 8$ (380~GHz) line of ortho-H$_2$O, from 50~K to above room temperature, also show a decreasing trend with decreasing temperature, with a minimum between 50 and 100~K. However, these PBx-sections are smaller than those calculated here with the PES by \cite{patk02}, as was also observed for the homologous sts-rates for inelastic collisions, as mentioned in Section 3.
It should be pointed out that the PBx-section $\sigma_{i f} (E_k)$ calculated by \cite{malu92} for the 183~GHz line does not show any resonance at low kinetic energy, as opposed to the present work where such resonances have been shown to contribute appreciably to the thermally averaged PBx-section $\sigma_{i f} (T)$.

We switch finally to the total PB-coefficients $\gamma_{i f}(T)$ between 20 and 120~K reported in Table~2.
They show an almost flat trend at higher temperatures, with a shallow minimum at $80 \leq T \leq 120$~K, and then grow up markedly for $T \leq 40$~K. While the thermal dependence of the PB-coefficients is usually modelled by empirical power laws like
\begin{equation}
  \gamma_{i f}(T)=   \gamma_{i f}(T_0) (T/T_0)^{\alpha},     \label{g-power-law}
\end{equation}
assuming a monotonic dependence with $T$,
in the present case such a simple dependence cannot be safely used because of the minima.
We propose, for the PB-coefficients of H$_2$O by helium, the alternative trend-law for the $ 20 \leq T \leq 120$~K range 
\begin{equation}
  \gamma_{i f}(T)=   g + a/T^2 + b \, (T - c)^2  .      \label{fit-law}
\end{equation}
The parameters $ g , a, b, c $, given in Table~2 for each line, provide an average accuracy better than $3 \%$ for most PB-coefficients with respect to the values reported in Table~2 for discrete temperatures, allowing for a safer interpolation than by using Eq.~(\ref{g-power-law}).

To conclude, the present study shows that the pressure broadening of H$_2$O spectral lines by He is a complex phenomenon, which does not seem amenable to simple relations with line frequency nor angular momentum.
Special care must be paid to the elastic contribution, which has not always been considered in the literature, but plays a significant role here: 
it is eventually responsible for the observed dispersion in the total PB-coefficients, can be dominant at 20~K, and is expected to further increase at lower temperatures.
The present results are thus intended as a useful guide for future laboratory measurements and astrophysical observations.

\section*{Acknowledgments}

Thanks are due to B. J. Drouin and J. C. Pearson for providing their experimental data on PB-coefficients for H$_2$O:He collisions, and to J. L. Domenech for helpful comments on the manuscript.
This work has been supported by the Spanish Ministerio de Econom\'{i}a y Competitividad (MINECO), grants FIS2013-48275-C2 and CONSOLIDER-ASTROMOL CSD2009-0038.

\appendix
\section{Appendix}

In this Appendix we present the details for the CC calculation with the MOLSCAT code of the 
PBx-section for a given electric-dipole transition, using the PES by \cite{patk02}.
The MOLSCAT code does not provide an option to just retrieve the elastic contribution of Eq.~(\ref{def-contrib-e}). 
However, it allows for the calculation of the {\em total} (elastic plus inelastic) PBx-section in a  straightforward manner.
Therefore, we have obtained the elastic contribution as the difference
\begin{equation}
 \sigma^{(el){\rm CC}}_{if}(T) = \sigma^{\rm CC}_{if}(T) -  \sigma^{(in){\rm CC}}_{if}(T) ,  \label{sigma CC-el}
\end{equation}
between the total PBx-section, $ \sigma^{\rm CC}_{if}(T)$, and the inelastic contribution, $\sigma^{(in){\rm CC}}_{if}(T)$.
\noindent

In the equation above, $\sigma^{\rm CC}_{if}(T)$ involves a thermal average of a set of PBx-sections in terms of the kinetic energy, $\sigma^{\rm CC}_{if}(E_k)$, according to the well-known Eq.~(20) of \cite{shaf73}. 
PBx-sections $\sigma^{\rm CC}_{if}(E_k)$ were computed using the hybrid log-derivative-Airy propagator by \cite{alex87} implemented in MOLSCAT, and are converged (with respect to basis set size, propagation parameters and number of partial waves) to better than 1\%. The calculations were carried out for kinetic energies up to 
850 cm$^{-1}$, taking care of using small energy steps  ($\Delta E_{k} \leq $ 0.5 cm$^{-1}$) for the lowest energies in order to describe properly the resonant structures of the cross sections. As discussed in the main text, such resonances play a relevant role in most PBx-sections at low temperature.
The resulting thermally averaged $\sigma^{\rm CC}_{if}(T)$ are converged within 5\% at 20~K, or better at higher temperatures.

On the other hand, the inelastic PBx-section $\sigma^{(in){\rm CC}}_{if}(T)$ in Eq.~(\ref{sigma CC-el})
can be computed either from a thermal average of the ordinary cross sections for inelastic collisions from the MOLSCAT output, or by means of  Eq.~(\ref{sigmaks}) using the sts-rates for inelastic collisions from the Patkowski PES reported by \cite{teje15} (P-rates therein); we checked that both procedures yield the same result.


\begin{deluxetable}{ccccccccccccrc}
\tabletypesize{\footnotesize}
\tablecaption{Pressure broadening cross sections $\sigma_{if}$, and coefficients $\gamma_{if}$, for rotational lines of H$_2$O.}
\tablehead{\multicolumn{6}{c}{ } & \multicolumn{2}{c}{inelastic}  & \multicolumn{2}{c}{total} & \multicolumn{4}{c}{fit} \\  \cline{7-8}  \cline{9-10}  \cline{11-14}
\colhead{o/p}  & \colhead{$i$} & \colhead{$f$} & \colhead{$\omega$} & \colhead{$\nu$ } &  \colhead{$T$} & \colhead{$\sigma_{if}$} & \colhead{$\gamma_{if}$} &  \colhead{$\sigma_{if}$} & \colhead{$\gamma_{if}$} & \colhead{$g$}   &  \colhead{$a$}  &  \colhead{$b \times 10^5$} &  \colhead{$c$}
}
\startdata
 para  & 5 & 6 &     183 &     6.1 &   20   &   6.58    &   1.817   & 19.29      & 5.330      & 1.120 & 1349.6 & 9.0712 & 115.89\\
       &   &   &         &         &   40   &   4.44    &   0.868   & 12.61      & 2.463      &       &        &        &       \\
       &   &   &         &         &   60   &   5.03    &   0.803   & 11.42      & 1.822      &       &        &        &       \\
       &   &   &         &         &   80   &   4.93    &   0.681   & 10.34      & 1.428      &       &        &        &       \\
       &   &   &         &         &  100   &   5.43    &   0.671   & 10.22      & 1.263      &       &        &        &       \\
       &   &   &         &         &  120   &   6.50    &   0.733   & 10.86      & 1.225      &       &        &        &       \\
 ortho & 7 & 8 &     380 &    12.7 &   20   &   6.06    &   1.674   & 13.21      & 3.649      & 1.051 & 884.60 & 4.8595 & 109.02\\
       &   &   &         &         &   40   &   4.89    &   0.955   &  9.31      & 1.820      &       &        &        &       \\
       &   &   &         &         &   60   &   5.64    &   0.900   &  9.08      & 1.448      &       &        &        &       \\
       &   &   &         &         &   80   &   5.80    &   0.802   &  8.71      & 1.204      &       &        &        &       \\
       &   &   &         &         &  100   &   6.68    &   0.826   &  9.27      & 1.146      &       &        &        &       \\
       &   &   &         &         &  120   &   7.56    &   0.853   &  9.94      & 1.121      &       &        &        &       \\
 ortho & 1 & 2 &     557 &    18.6 &   20   &   1.93    &   0.534   &  6.85      & 1.893      & 0.872 & 393.59 & 2.2416 & 60.12 \\
       &   &   &         &         &   40   &   2.73    &   0.534   &  5.69      & 1.112      &       &        &        &       \\
       &   &   &         &         &   60   &   4.12    &   0.657   &  6.36      & 1.015      &       &        &        &       \\
       &   &   &         &         &   80   &   4.75    &   0.656   &  6.62      & 0.914      &       &        &        &       \\
       &   &   &         &         &  100   &   6.07    &   0.750   &  7.70      & 0.952      &       &        &        &       \\
       &   &   &         &         &  120   &   7.23    &   0.816   &  8.70      & 0.981      &       &        &        &       \\
 para  & 3 & 4 &     752 &    25.1 &   20   &   6.28    &   1.736   & 10.42      & 2.879      & 0.891 & 703.16 & 3.6962 & 98.34 \\
       &   &   &         &         &   40   &   4.65    &   0.909   &  7.29      & 1.425      &       &        &        &       \\
       &   &   &         &         &   60   &   5.45    &   0.870   &  7.52      & 1.200      &       &        &        &       \\
       &   &   &         &         &   80   &   5.42    &   0.748   &  7.18      & 0.992      &       &        &        &       \\
       &   &   &         &         &  100   &   6.00    &   0.741   &  7.58      & 0.936      &       &        &        &       \\
       &   &   &         &         &  120   &   7.17    &   0.809   &  8.63      & 0.973      &       &        &        &       \\
 para  & 2 & 3 &     988 &    33.0 &   20   &   4.52    &   1.249   &  7.91      & 2.186      & 0.868 & 496.42 & 2.2328 & 77.48 \\
       &   &   &         &         &   40   &   3.84    &   0.750   &  6.03      & 1.179      &       &        &        &       \\
       &   &   &         &         &   60   &   4.96    &   0.791   &  6.71      & 1.070      &       &        &        &       \\
       &   &   &         &         &   80   &   5.19    &   0.717   &  6.71      & 0.927      &       &        &        &       \\
       &   &   &         &         &  100   &   5.94    &   0.734   &  7.30      & 0.902      &       &        &        &       \\
       &   &   &         &         &  120   &   7.25    &   0.817   &  8.50      & 0.959      &       &        &        &       \\
 ortho & 5 & 6 &    1097 &    36.6 &   20   &   5.60    &   1.548   &  9.13      & 2.523      & 0.934 & 595.92 & 2.1550 & 87.37 \\
       &   &   &         &         &   40   &   4.76    &   0.930   &  6.83      & 1.335      &       &        &        &       \\
       &   &   &         &         &   60   &   5.66    &   0.903   &  7.25      & 1.157      &       &        &        &       \\
       &   &   &         &         &   80   &   5.89    &   0.813   &  7.24      & 1.000      &       &        &        &       \\
       &   &   &         &         &  100   &   6.84    &   0.845   &  8.05      & 0.995      &       &        &        &       \\
       &   &   &         &         &  120   &   7.77    &   0.876   &  8.89      & 1.003      &       &        &        &       \\
 para  & 1 & 2 &    1113 &    37.1 &   20   &   2.52    &   0.697   &  2.91      & 0.805      & 0.595 &  79.01 & 1.6770 &  0.00 \\
       &   &   &         &         &   40   &   3.00    &   0.587   &  3.21      & 0.627      &       &        &        &       \\
       &   &   &         &         &   60   &   4.48    &   0.715   &  4.62      & 0.736      &       &        &        &       \\
       &   &   &         &         &   80   &   5.03    &   0.695   &  5.13      & 0.709      &       &        &        &       \\
       &   &   &         &         &  100   &   5.98    &   0.739   &  6.06      & 0.749      &       &        &        &       \\
       &   &   &         &         &  120   &   7.48    &   0.843   &  7.53      & 0.850      &       &        &        &       \\
 ortho & 4 & 6 &    1153 &    38.5 &   20   &   5.41    &   1.496   &  9.29      & 2.568      & 0.958 & 588.76 & 2.4599 & 94.36 \\
       &   &   &         &         &   40   &   4.67    &   0.912   &  7.07      & 1.381      &       &        &        &       \\
       &   &   &         &         &   60   &   5.54    &   0.883   &  7.46      & 1.190      &       &        &        &       \\
       &   &   &         &         &   80   &   5.77    &   0.797   &  7.45      & 1.029      &       &        &        &       \\
       &   &   &         &         &  100   &   6.70    &   0.828   &  8.23      & 1.017      &       &        &        &       \\
       &   &   &         &         &  120   &   7.62    &   0.859   &  9.04      & 1.020      &       &        &        &       \\
 ortho & 6 & 7 &    1163 &    38.8 &   20   &   5.77    &   1.595   &  6.90      & 1.907      & 0.828 & 428.22 & 1.0157 & 45.56 \\
       &   &   &         &         &   40   &   4.87    &   0.952   &  5.50      & 1.075      &       &        &        &       \\
       &   &   &         &         &   60   &   5.71    &   0.911   &  6.22      & 0.993      &       &        &        &       \\
       &   &   &         &         &   80   &   5.90    &   0.815   &  6.36      & 0.879      &       &        &        &       \\
       &   &   &         &         &  100   &   6.81    &   0.842   &  7.26      & 0.897      &       &        &        &       \\
       &   &   &         &         &  120   &   7.71    &   0.870   &  8.16      & 0.920      &       &        &        &       \\
 para  & 4 & 5 &    1229 &    41.0 &   20   &   7.02    &   1.940   & 10.45      & 2.888      & 0.842 & 733.26 & 3.3809 & 98.75 \\
       &   &   &         &         &   40   &   4.86    &   0.949   &  7.09      & 1.385      &       &        &        &       \\
       &   &   &         &         &   60   &   5.48    &   0.874   &  7.26      & 1.157      &       &        &        &       \\
       &   &   &         &         &   80   &   5.32    &   0.735   &  6.85      & 0.946      &       &        &        &       \\
       &   &   &         &         &  100   &   5.83    &   0.720   &  7.20      & 0.890      &       &        &        &       \\
       &   &   &         &         &  120   &   6.93    &   0.782   &  8.19      & 0.924      &       &        &        &       \\
 ortho & 3 & 4 &    1661 &    55.4 &   20   &   6.18    &   1.708   & 14.12      & 3.900      & 1.157 & 910.21 & 5.6451 & 110.87\\
       &   &   &         &         &   40   &   5.11    &   0.998   & 10.21      & 1.994      &       &        &        &       \\
       &   &   &         &         &   60   &   5.98    &   0.953   &  9.97      & 1.591      &       &        &        &       \\
       &   &   &         &         &   80   &   6.19    &   0.855   &  9.59      & 1.325      &       &        &        &       \\
       &   &   &         &         &  100   &   7.15    &   0.883   & 10.19      & 1.259      &       &        &        &       \\
       &   &   &         &         &  120   &   8.09    &   0.913   & 10.88      & 1.227      &       &        &        &       \\
 ortho & 1 & 3 &    1670 &    55.7 &   20   &   3.90    &   1.077   &  7.00      & 1.935      & 0.898 & 408.53 & 1.5086 & 50.59 \\
       &   &   &         &         &   40   &   3.84    &   0.751   &  5.82      & 1.137      &       &        &        &       \\
       &   &   &         &         &   60   &   5.04    &   0.804   &  6.59      & 1.052      &       &        &        &       \\
       &   &   &         &         &   80   &   5.53    &   0.764   &  6.85      & 0.947      &       &        &        &       \\
       &   &   &         &         &  100   &   6.70    &   0.828   &  7.90      & 0.976      &       &        &        &       \\
       &   &   &         &         &  120   &   7.78    &   0.878   &  8.89      & 1.003      &       &        &        &       \\
 ortho & 3 & 5 &    1717 &    57.3 &   20   &   6.37    &   1.760   &  7.64      & 2.110      & 0.910 & 473.19 & 1.1446 & 56.06 \\
       &   &   &         &         &   40   &   5.20    &   1.017   &  6.08      & 1.187      &       &        &        &       \\
       &   &   &         &         &   60   &   6.10    &   0.973   &  6.82      & 1.088      &       &        &        &       \\
       &   &   &         &         &   80   &   6.31    &   0.872   &  6.96      & 0.961      &       &        &        &       \\
       &   &   &         &         &  100   &   7.29    &   0.901   &  7.90      & 0.976      &       &        &        &       \\
       &   &   &         &         &  120   &   8.24    &   0.930   &  8.83      & 0.996      &       &        &        &       \\
 para  & 6 & 7 &    1919 &    64.0 &   20   &   6.86    &   1.894   & 11.17      & 3.085      & 0.846 & 808.88 & 3.4238 & 98.97 \\
       &   &   &         &         &   40   &   4.62    &   0.903   &  7.37      & 1.440      &       &        &        &       \\
       &   &   &         &         &   60   &   5.23    &   0.834   &  7.40      & 1.181      &       &        &        &       \\
       &   &   &         &         &   80   &   5.12    &   0.708   &  6.98      & 0.964      &       &        &        &       \\
       &   &   &         &         &  100   &   5.64    &   0.697   &  7.30      & 0.902      &       &        &        &       \\
       &   &   &         &         &  120   &   6.74    &   0.760   &  8.27      & 0.933      &       &        &        &       \\
 para  & 3 & 6 &    2164 &    72.2 &   20   &   5.84    &   1.613   &  7.33      & 2.026      & 0.731 & 498.66 & 1.7987 & 69.96 \\
       &   &   &         &         &   40   &   4.24    &   0.828   &  5.26      & 1.028      &       &        &        &       \\
       &   &   &         &         &   60   &   5.01    &   0.799   &  5.83      & 0.930      &       &        &        &       \\
       &   &   &         &         &   80   &   5.02    &   0.694   &  5.72      & 0.791      &       &        &        &       \\
       &   &   &         &         &  100   &   5.60    &   0.692   &  6.24      & 0.771      &       &        &        &       \\
       &   &   &         &         &  120   &   6.74    &   0.761   &  7.33      & 0.827      &       &        &        &       \\
 para  & 6 & 8 &    2392 &    79.8 &   20   &   6.48    &   1.792   &  7.39      & 2.043      & 0.670 & 539.11 & 1.4140 & 59.83 \\
       &   &   &         &         &   40   &   4.41    &   0.861   &  5.01      & 0.980      &       &        &        &       \\
       &   &   &         &         &   60   &   5.02    &   0.801   &  5.51      & 0.879      &       &        &        &       \\
       &   &   &         &         &   80   &   4.93    &   0.681   &  5.36      & 0.740      &       &        &        &       \\
       &   &   &         &         &  100   &   5.42    &   0.670   &  5.81      & 0.718      &       &        &        &       \\
       &   &   &         &         &  120   &   6.50    &   0.734   &  6.87      & 0.775      &       &        &        &       \\
 ortho & 5 & 8 &    2640 &    88.1 &   20   &   5.88    &   1.626   &  6.77      & 1.872      & 0.795 & 429.17 & 0.9270 & 35.43 \\
       &   &   &         &         &   40   &   4.77    &   0.933   &  5.34      & 1.043      &       &        &        &       \\
       &   &   &         &         &   60   &   5.59    &   0.891   &  6.04      & 0.963      &       &        &        &       \\
       &   &   &         &         &   80   &   5.79    &   0.800   &  6.18      & 0.853      &       &        &        &       \\
       &   &   &         &         &  100   &   6.71    &   0.829   &  7.06      & 0.873      &       &        &        &       \\
       &   &   &         &         &  120   &   7.62    &   0.859   &  7.95      & 0.897      &       &        &        &       \\
 ortho & 2 & 4 &    2774 &    92.5 &   20   &   4.22    &   1.165   &  7.63      & 2.107      & 0.900 & 453.09 & 2.2997 & 76.40 \\
       &   &   &         &         &   40   &   4.00    &   0.781   &  6.12      & 1.196      &       &        &        &       \\
       &   &   &         &         &   60   &   5.05    &   0.806   &  6.71      & 1.070      &       &        &        &       \\
       &   &   &         &         &   80   &   5.41    &   0.747   &  6.82      & 0.942      &       &        &        &       \\
       &   &   &         &         &  100   &   6.52    &   0.805   &  7.77      & 0.960      &       &        &        &       \\
       &   &   &         &         &  120   &   7.54    &   0.850   &  8.67      & 0.978      &       &        &        &       \\
 para  & 2 & 5 &    2969 &    99.0 &   20   &   5.26    &   1.453   &  9.79      & 2.704      & 0.858 & 675.83 & 3.1242 & 90.08 \\
       &   &   &         &         &   40   &   4.05    &   0.791   &  6.80      & 1.329      &       &        &        &       \\
       &   &   &         &         &   60   &   4.98    &   0.794   &  7.08      & 1.130      &       &        &        &       \\
       &   &   &         &         &   80   &   5.10    &   0.704   &  6.86      & 0.947      &       &        &        &       \\
       &   &   &         &         &  100   &   5.77    &   0.713   &  7.32      & 0.904      &       &        &        &       \\
       &   &   &         &         &  120   &   7.01    &   0.790   &  8.41      & 0.948      &       &        &        &       \\
 para  & 4 & 7 &    3331 &   111.1 &   20   &   7.30    &   2.017   &  7.67      & 2.120      & 0.725 & 546.10 & 1.5094 & 61.76 \\
       &   &   &         &         &   40   &   5.03    &   0.984   &  5.31      & 1.038      &       &        &        &       \\
       &   &   &         &         &   60   &   5.68    &   0.905   &  5.91      & 0.943      &       &        &        &       \\
       &   &   &         &         &   80   &   5.52    &   0.762   &  5.73      & 0.792      &       &        &        &       \\
       &   &   &         &         &  100   &   6.04    &   0.746   &  6.24      & 0.772      &       &        &        &       \\
       &   &   &         &         &  120   &   7.17    &   0.808   &  7.37      & 0.832      &       &        &        &       \\
 ortho & 3 & 7 &    3977 &   132.7 &   20   &   6.54    &   1.808   & 11.57      & 3.196      & 1.051 & 758.84 & 3.5965 & 102.68\\
       &   &   &         &         &   40   &   5.32    &   1.039   &  8.44      & 1.648      &       &        &        &       \\
       &   &   &         &         &   60   &   6.15    &   0.982   &  8.58      & 1.369      &       &        &        &       \\
       &   &   &         &         &   80   &   6.32    &   0.873   &  8.39      & 1.159      &       &        &        &       \\
       &   &   &         &         &  100   &   7.26    &   0.897   &  9.13      & 1.128      &       &        &        &       \\
       &   &   &         &         &  120   &   8.18    &   0.923   &  9.92      & 1.119      &       &        &        &       \enddata
\tablecomments{Units: Line frequency $\omega$ in GHz, wavenumber $\nu$ in cm$^{-1}$, $T$ in K, $\sigma_{if}$ in  \AA$^2$,  $\gamma_{if}$ in MHz/Torr (divide by 39.45 to translate into cm$^{-1}$/atm).
 Fit parameters $g, a, b, c$ according to Eq.~(\ref{fit-law}).}
\end{deluxetable}

\end{document}